# MODELING OF NUCLEAR REACTIONS WITH LANGEVIN CALCULATIONS

S. Amano*, Y. Aritomo, Y. Miyamoto, S. Ishizaki and M. Okubayashi

*Faculty of Science and Engineering Kindai University, Higashi-Osaka, 577-8502 Japan*

*E-mail: amano_shota@kindai.ac.jp*

**Abstract**—The mass angle distribution shows a strong correlation between mass and angle when quasifission events are dominant. Therefore, as long as quasifission events are dominant, the mass angle distribution is characterized in that diagonal correlation appears. This diagonal correlation could not be reproduced in our previous model that is before introducing $f_{ina}$ and $\gamma_t^0$ model parameters. In this study, we clarify the indeterminate parameters included in the model to reproduce the diagonal correlation appearing in the mass angle distribution in the $^{48}$Ti + $^{186}$W reaction system. As a result, $f_{ina}$ and $\gamma_t^0$ of model parameters found to be key parameters for MAD. it was also found that the balance between $f_{ina}$ and $\gamma_t^0$ parameters values is important for the strong correlation between mass and angle.

## INTRODUCTION

Production of neutron-rich nuclei is very important in research on the origin of elements in the universe and chemical evolution of the universe. However, it has become difficult to produce new neutron-rich nuclei and superheavy nuclei by heavy ion fusion reactions. Therefore, multinucleon transfer reactions has been proposed to produce new neutron-rich nuclei and superheavy nuclei [1,2,3]. JAEA (Japan Atomic Energy Agency) plans to conduct a unique experiment using 99th element Es. They are trying to reach a new area of the nuclear chart by using multi-nucleon transfer reactions with Es. Thus, the nucleon transfer reaction is becoming the mainstream production method in future heavy ion nuclear reactions. In the experiment related to the nucleon transfer reaction, the Australian National University group measured mass angle distribution to analyze the fusion mechanism [4]. From these results, it became clear that there is a correlation between the fission fragment mass and the emission angle, and the mass angle distribution is characterized by the incident channel. The purpose of this study is to improve the reproducibility of the diagonal correlation, which is a characteristic of the mass angle distribution. For this purpose, we developed the dynamical model to apply the nucleon transfer reaction. Specifically, We change the indeterminate parameters included in our model to investigate the parameter dependence of the mass angle distribution and the mass distribution. We then clarify the indeterminate parameters by comparing the calculation results with the experimental results.

## MODEL

### *Potential energy surface*

The initial stage of the nucleon transfer reactions consists of two parts: (1) the system is placed in the ground state of the projectile and target because the reaction proceeds is too fast for nucleons to reconfigure a single particle state (2) The part where the system relaxes to the ground state of the entire composite system which changes the potential energy surface to an adiabatic one. Therefore, we consider the time evolution of potential energy from the diabatic one $V_{\text{diab}}(q)$ to adiabatic one $V_{\text{adiab}}(q)$. Here, $q$ denotes a set of collective coordinates representing nuclear deformation. The diabatic potential is calculated by a folding procedure using effective nucleon-nucleon interaction [5-7]. As a characteristic of the diabatic potential, "potential wall" appears due to the Overlap region of collision system which corresponds to the hard core representing the incompressibility of nuclear material. However, the adiabatic potential energy of the system is calculated using an extended two-center shell model [7]. Then, we connect the diabatic and the adiabatic potentials with a time-dependent weighting function as follows:

$$V = V_{\text{diab}}(q)f(t) + V_{\text{adiab}}(q)[1 - f(t)],$$

$$f(t) = \exp\left(-\frac{t}{\tau}\right). \tag{1}$$

Where $t$ is the interaction time and $f(t)$ is the weighting function included the relaxation time $\tau$. We use the relaxation time $\tau = 10^{-21}$s proposed in [8-10]. We use the two-center parameterization [11,12] as coordinates to represent nuclear deformation.

To solve the dynamical equation numerically and avoid the huge computation time, we strictly limited the number of degrees of freedom and employ three parameters as follows: $z_0$ (distance between the centers of two potentials), $\delta$ (deformation of fragment), and $\alpha$ (mass asymmetry of colliding nuclei); $\alpha = (A_1 - A_2)/(A_1 + A_2)$, where $A_1$ and $A_2$ not only stand for the mass numbers of the target and projectile, respectively [5,13] but also are then used to indicate mass numbers of the two fission fragments. The mass ratio $M_R$ is defined as $M_R = A_1/(A_1 + A_2)$. As shown in Fig. 1 in Ref. [11], the parameter $\delta$ is defined as $\delta = 3(a-b)/(2a+b)$, where $a$ and $b$ represent the half length of the ellipse axes in the $z_0$ and $\rho$ directions, respectively. We assume that each fragment has the same deformation as a first step. In addition, we use scaling to save computation time and use the coordinate $z$ defined as $z = z_0/(R_{CN}B)$, where $R_{CN}$ denotes the radius of the spherical compound nucleus and the parameter $B$ is defined as $B = (3+\delta)/(3-2\delta)$.

*Dynamical equations*

Using the Langevin equation, we perform trajectory calculations of the time-dependent unified potential energy [5,6,13]. We start trajectory calculations from a sufficiently long distance between both nuclei [13]. So, we take into account the nucleon transfer for slightly separated nuclei [5]. Such intermediate nucleon exchange also plays an important role in the fusion process near and below the Coulomb barrier. Process for the separated nucleon transfer use the procedure described in Refs. [5,6]:

$$\frac{d\alpha}{dt} = \frac{2}{A_{CN}} D_A^{(1)}(\alpha) + \frac{2}{A_{CN}} \sqrt{D_A^{(2)}(\alpha)} \gamma_\alpha(t) \qquad (2)$$

This is the Langevin equation ignoring the inertial mass for the mass asymmetry parameter and represents the change in the asymmetric parameter $\alpha$ due to the drift (first term on the right) and diffusion (second term on the right) processes. It is obtained by a specific approximation starting from the master equation for the different particle-hole state transitions. Such a master equation that gives the discrete change in the number of nucleons is transformed from the Fokker-Planck equation to the above-mentioned Langevin equation, and finally transformed to the equation of the continuous variable $\alpha$ [5,6]. Although *N/Z* equilibrium affects the transfer process, our present model doesn't consider *N/Z* equilibrium. Thus, we assume that *N/Z* equilibrium is achieved quickly as a first approximation using the same model in reference [5]. After the window of the contact nuclei is sufficiently opened (so-called "mononucleus state"), the evolution process of the mass asymmetry parameter $\alpha$ switches from Eq. (2) to Langevin equation according to the procedure described in Ref. [13].

The multidimensional Langevin equation [5,13,14] was unified as follows:

$$\frac{dq_i}{dt} = (m^{-1})_{ij} p_j,$$

$$\frac{dp_i}{dt} = -\frac{\partial V}{\partial q_i} - \frac{1}{2}\frac{\partial}{\partial q_i}(m^{-1})_{jk} p_j p_k - \gamma_{ij}(m^{-1})_{jk} p_k + g_{ij} R_j(t),$$

$$\frac{d\theta}{dt} = \frac{\ell}{\mu_R R^2}, \quad \frac{d\varphi_1}{dt} = \frac{L_1}{\Im_1}, \quad \frac{d\varphi_2}{dt} = \frac{L_2}{\Im_2},$$

$$\frac{d\ell}{dt} = -\frac{\partial V}{\partial \theta} - \gamma_{tan}\left(\frac{\ell}{\mu_R R^2} - \frac{L_1}{\Im_1}a_1 - \frac{L_2}{\Im_2}a_2\right) R + R g_{tan} R_{tan}(t),$$

$$\frac{dL_1}{dt} = -\frac{\partial V}{\partial \varphi_1} - \gamma_{tan}\left(\frac{\ell}{\mu_R R^2} - \frac{L_1}{\Im_1}a_1 - \frac{L_2}{\Im_2}a_2\right) a_1 - a_1 g_{tan} R_{tan}(t),$$

$$\frac{dL_2}{dt} = -\frac{\partial V}{\partial \varphi_2} + \gamma_{tan}\left(\frac{\ell}{\mu_R R^2} - \frac{L_1}{\Im_1}a_1 - \frac{L_2}{\Im_2}a_2\right) a_2 - a_2 g_{tan} R_{tan}(t). \quad (3)$$

The collective coordinates $q_i$ represent $z$, $\delta$, and $\alpha$, the symbol $p_i$ denotes momentum conjugated to $q_i$, and $V$ is the multidimensional potential energy. The definition of other parameters is shown in Fig. 2 in [27]: The symbols $\theta$ and $\ell$ are the relative orientation of nuclei and relative angular momentum respectively. $\varphi_1$ and $\varphi_2$ stand for the rotation angles of the nuclei in the reaction plane (their moment of inertia and angular momenta are $\mathfrak{I}_{1,2}$ and $L_{1,2}$, respectively), $a_{1,2} = R/2 \pm (R_1 - R_2)/2$ is the distance from the center of the fragment to the middle point between the nuclear surfaces, and $R_{1,2}$ is the nuclear radii. The symbol $R$ is distance between the nuclear centers. The total angular momentum $L = \ell + L_1 + L_2$ is preserved. The symbol $\mu_R$ is reduced mass, and $\gamma_{\tan}$ is the tangential friction force of the colliding nuclei. Here, it is called sliding friction. The phenomenological nuclear friction forces for separated nuclei are expressed in terms of $\gamma_R^F$ and $\gamma_{\tan}^F$ for radial and sliding friction using the Woods-Saxon radial form factor described in Refs. [5,6]. The Radial and sliding friction are described as $\gamma_R^F = \gamma_R^0 F(\zeta)$, $\gamma_{\tan} = \gamma_t^0 F(\zeta)$, where the radial form factor $F(\zeta) = \left(1 + e^\zeta\right)^{-1}$, $\zeta = (\xi - \rho_F)/a_F$. $\gamma_R^0$ and $\gamma_t^0$ denote the strength of the radial and tangential frictions, respectively. $\rho_F \sim 2$ fm and $a_F \sim 0.6$ fm are the model parameters, and $\xi$ is the distance between the nuclear surfaces $\xi = R - R_{\text{contact}}$, where $R_{\text{contact}} = R_1 + R_2$ [5]. The symbols separated by $m_{ij}$ and $\gamma_{ij}$ stand foer the shape-dependent collective inertia and friction tensors elements, respectively. For separated nuclei, the reduced mass and phenomenological friction forces $\gamma_R^F$ are used.

Then we switch the phenomenological friction to that of a mononuclear system with a smoothing function $\theta(\xi) = \left(1 + \exp^{-\xi/0.3}\right)^{-1}$ [5,6]. For the mononuclear system, the wall and window one-body dissipation $\gamma_R^{\text{one}}$ is adopted for the friction tensor [15–22]. $\gamma_{zz}$, $\gamma_{\delta\delta}$, and $\gamma_{\alpha\alpha}$ denote the absolute values of the wall and window dissipation for the radial direction, the deformation, and the mass asymmetry. The resulting radial friction is expressed as $\gamma_R = \gamma_R^{\text{one}} + \theta(\xi)\gamma_R^F(\xi - \rho)$ which is related to $\gamma_{zz}$. We adopted the hydrodynamic inertia tensor $m_{ij}$ in the Werner-Wheeler approximation for the velocity field [23]. The normalized random force $R_i(t)$ is assumed to be white noise: $\langle R_i(t)\rangle = 0$ and $\langle R_i(t_1)R_j(t_2)\rangle = 2\delta_{ij}\delta(t_1 - t_2)$. According to the Einstein relation, the strength of the random force $g_{ij}$ is given $\gamma_{ij}T = \sum_k g_{ij}g_{jk}$, where $T$ is the temperature of the compound nucleus calculated from the intrinsic energy of the composite system.

The adiabatic potential energy is defined as

$$V_{\text{adiab}}(q, L, T) = V_{\text{LD}}(q) + \frac{\hbar^2 L(L+1)}{2I(q)} + V_{\text{SH}}(q, T)$$

$$V_{\text{LD}}(q) = E_S(q) + E_C(q),$$

$$V_{\text{SH}}(q, T) = E_{\text{shell}}^0(q)\Phi(T),$$

$$\Phi(T) = \exp\left(-\frac{E^*}{E_d}\right),$$

Here, $I(q)$ represents the moment of inertia of the rigid body with deformation $q$. The centrifugal energy generated from the angular momentum $L$ of the rigid body is also taken into account. $V_{\text{LD}}$ and $V_{\text{SH}}$ are the potential energy of the finite range liquid drop model and the shell correction energy that takes into account temperature dependence, respectively. The symbol $E_{\text{shell}}^0$ indicates the shell correction energy at $T = 0$. The temperature dependence factor $\Phi(T)$ is explained in Ref. [24], where $E^*$ indicates the excitation energy of the compound nucleus. $E^*$ is given $E^* = aT^2$, where $a$ is the level density parameter. The shell damping energy $E_d$ is selected as 20 MeV. This value is given by Ignatyuk et al. [25]. The symbols $E_S$ and $E_C$ stand for generalized surface energy [26] and Coulomb energy, respectively.

## RESULTS

To investigate the dependence of unknown parameters in the model, we changed two parameters and calculated the mass angle distribution and mass distribution in the reaction $^{48}$Ti + $^{186}$W at E$_{\text{cm}}$ = 187.87MeV. Then, we compared them with the experimental results [4]. We changed the moment of inertia parameter $I' = f_{ina} \cdot I$. Another one is the strength of the tangential friction $\gamma_t^0$.

### Mass angle distribution

Figure. 1 shows the calculation results of mass angle distribution in the reaction $^{48}$Ti + $^{186}$W with several values of $f_{ina}$ and $\gamma_t^0$. The experimental result is shown in Fig. 2(a). When $f_{ina}$ is large, the rotational energy is small, so the nucleus become difficult to move.

That is, after the projectile nucleus comes into contact with the target nucleus, the projectile nucleus tends to move around ($\theta_{cm} = 180° \rightarrow 90°$) the target nucleus.

Conversely, when we choose small $f_{ina}$, the projectile nucleus tends to move around ($\theta_{cm} = 180° \rightarrow 90° \rightarrow 0° \rightarrow 90° \rightarrow 180°$) the target nucleus.

When $\gamma_t^0$ is large, the correlation between mass and angle disappears. The results in Fig. 1 ($f_{ina} = 1.5$, $\gamma_t^0 = 0.5$ and $f_{ina} = 1.5$, $\gamma_t^0 = 1.0$) reproduce the experimental result well.

*Mass distribution*

Next, we discuss the calculation results of mass distribution. Fig. 3 show the calculation results of the mass distribution of $^{48}$Ti + $^{186}$W when $f_{ina}$ and $\gamma_t^0$ are changed, respectively. The experimental result is shown in Fig. 2(b). Even though we changed $f_{ina}$, there is no significant change in the fission fragment mass. when $\gamma_t^0$ becomes larger, mass symmetric fission tends to increase. Comparing experimental results with calculation results in mass distributions for the reaction $^{48}$Ti + $^{186}$W, we found $\gamma_t^0 = 0.5$ or $1.0$ to reproduce the experimental result well.

As a consequence of the above, we determined (1) $f_{ina} = 1.5$, $\gamma_t^0 = 0.5$ and (2) $f_{ina} = 1.5$, $\gamma_t^0 = 1.0$ as the reasonable values of parameters to reproduce the diagonal correlation of mass angle distribution in the $^{48}$Ti + $^{186}$W reaction system.

CONCLUSIONS

Nucleon transfer reactions is becoming the mainstream production method in future heavy ion nuclear reactions. Nevertheless, a quantitative prediction model for the production rate has not been established. The reason is nucleon transfer reactions are a very complex problem. Therefore, in this study, we clarified the indeterminate model parameters in the $^{48}$Ti + $^{186}$W reaction system. We also reproduced the diagonal correlation of mass angle distribution in the $^{48}$Ti + $^{186}$W reaction system within our model. However, what about other reaction systems? We need to systematically evaluate $f_{ina}$ and $\gamma_t^0$ of model parameters in the future. Understanding nuclear reactions and structure, it is necessary to take very long research. We believe that this steady effort will lead to the understanding for accurately calculating the cross section of unknown nuclei by theoretical calculations.

## REFERENCES

1. V. I. Zagrebaev and W. Greiner, Phys. Rev. Lett., 2008, vol. 101, p. 122701.
2. A.V. Karpov and V. V. Saiko, Phys. Rev. C, 2017, vol. 96, p. 024618.
3. V. I. Zagrebaev, Yu. Ts. Oganessian, M. G. Itkis and W. Greiner, Phys. Rev. C, 2008, vol. 73, p. 122701.
4. R. du Riez, E. Williams, D. J. Hinde, M. Dasgupta, M. Evers, C. J. Lin, D. H. Luong, C. Simenel, and A. Wakhle, Phys. Rev. C, 2013, vol. 88, p. 054618.
5. V. Zagrebaev and W.Greiner, J. Phys. G, 2005, vol. 31, p. 825.
6. V. Zagrebaev and W.Greiner, J. Phys. G, 2007, vol. 34, p. 1 ; 2007, vol. 34, p. 2265.
7. V. I. Zagrebaev, A. V. Karpov, Y. Aritomo, M. A. Naumenko, and W. Greiner, Phys. Part. Nuclei, 2007, vol. 38, p. 469 ; NRV codes for driving potential[http://nrv.jinr.ru/nrv/].
8. G.F. Bertsch, Z. Phys. A, 1978, vol. 289, p. 103.
9. W. Cassing and W. Nörenberg, Nucl. Phys. A, 1983, vol. 401, p. 467.
10. A. Diaz-Torres, Phys. Rev. C, 2004, vol. 69, p. 021603.

11. J. Maruhn and W. Greiner, Z. Phys, 1972, vol. 251, p. 431.
12. K. Sato, A. Iwamoto, K. Harada, S. Yamaji and S. Yoshida, Z. Phys. A, 1978, vol. 288, p. 383.
13. Y. Aritomo and M. Ohta, Nucl. Phys. A, 2004, vol. 744, p. 3.
14. Y. Aritomo, Phys. Rev. C, 2009, vol. 80, p. 064604.
15. J. Blocki, Y. Boneh, J. R. Nix, J. Randrup, M. Robel, A. J. Sierk and W. J. Swiatecki, Ann. Phys, 1978, vol. 113, p. 330.
16. J. R. Nix and A. J. Sierk, Nucl. Phys. A, 1984, vol. 428, p. 161c.
17. J. Randrup and W. J. Swiatecki, Nucl. Phys. A, 1984, vol. 429, p. 105.
18. H. Feldmeier, Rep. Prog. Phys, 1987, vol. 50, p. 915.
19. N. Carjan, A. J. Sierk, and J. R. Nix, Nucl. Phys. A, 1986, vol. 452, p. 381.
20. N. Carjan, T. Wada and Y.Abe, in *Towards a Unified Picture of Nuclear Dynamics*, AIP Conference Proceeding No. 250 (AIP, New York, 1992).
21. T. Wada, Y, Abe and N. Carjan, Phys. Rev. Lett, 1993, vol. 70, p. 3538.
22. T. Asano, T. Wada, M. Ohta, S. Yamaji and H. Nakahara, J. Nucl. Radio Sci,
    2006, vol. 7, p. 7.
23. K. T. R. Davies, A. J. Sierk and J. R. Nix, Phys. Rev. C, 1976, vol. 13, p. 2385.
24. Y. Aritomo, Nucl. Phys. A, 2006, vol. 780, p. 222.
25. A. N. Ignatyuk, G. N. Smirenkin and A. S Tishin, Sov. J. Nucl. Phys, 1975,
    vol. 21, p. 255.
26. H. J. Krappe, J. R. Nix and A. J. Sierk, Phys. Rev. C, 1979, vol. 20, p. 992.
27. Y. Aritomo, S. Chiba, and K. Nishio, Phys. Rev. C, 2011, vol. 84, p. 024602.

FIGURE CAPTIONS

For the manuscript

S. Amano, Y. Aritomo, Y. Miyamoto, S. Ishizaki and M. Okubayashi

FUSION PROCESS IN TRANSFER REACTIONS

WITH LANGEVIN CALCULATION

**Fig. 1.** Parameter dependence calculation result of mass angle distribution in the reaction $^{48}$Ti + $^{186}$W. The vertical and horizontal changes represent $\gamma_t^0$ and $f_{ina}$, respectively.

**Fig. 2.** (a) The experimental result of mass angle distribution in the reaction $^{48}$Ti + $^{186}$W [4]. (b) The experimental result of mass distribution in the reaction $^{48}$Ti + $^{186}$W [4].

**Fig. 3.** Parameter dependence calculation result of mass distribution in the reaction $^{48}$Ti + $^{186}$W. The vertical and horizontal changes represent $\gamma_t^0$ and $f_{ina}$, respectively.

Fig. 1 for manuscript "FUSION PROCESS IN TRANSFER REACTION WITH…"

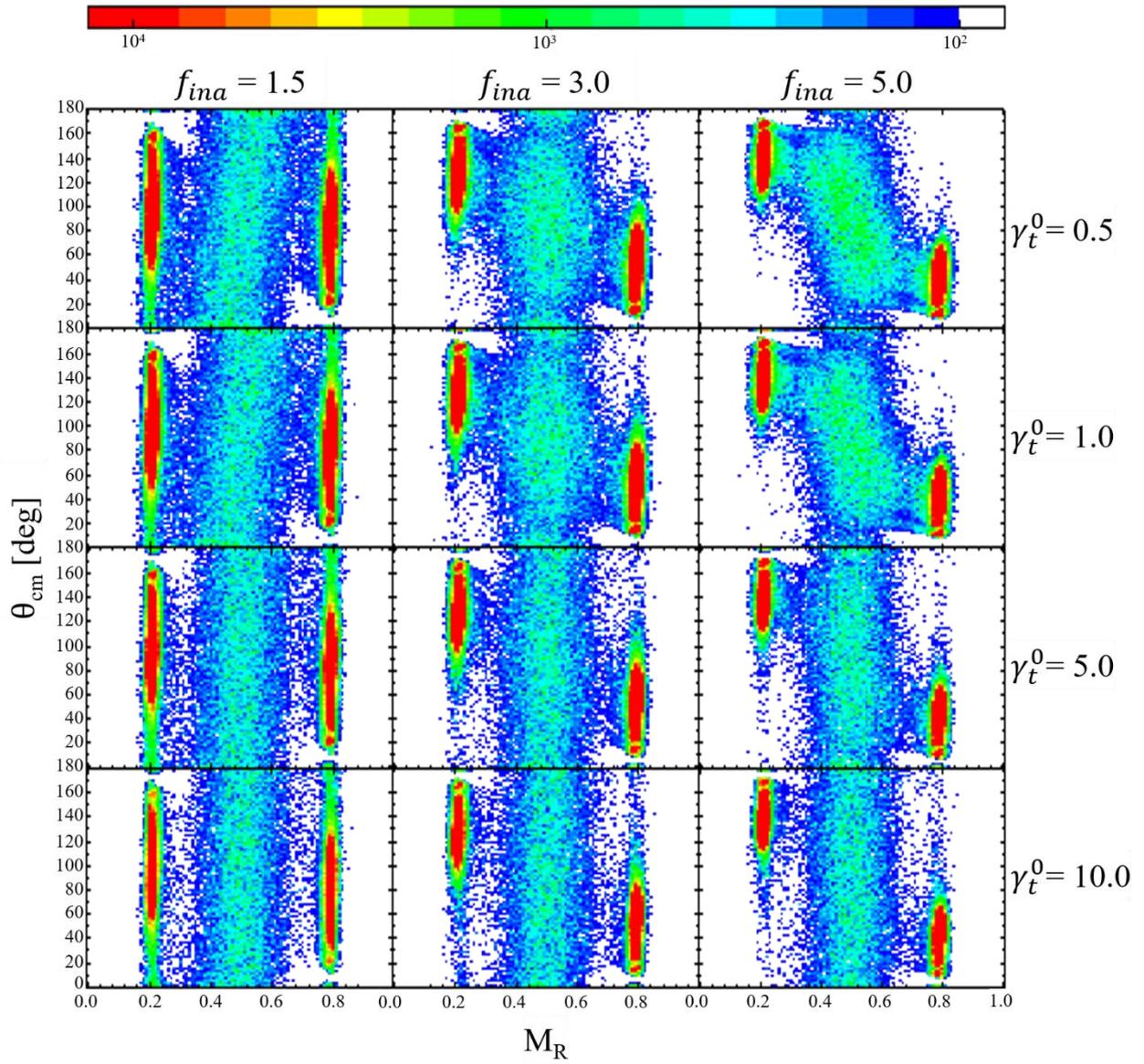

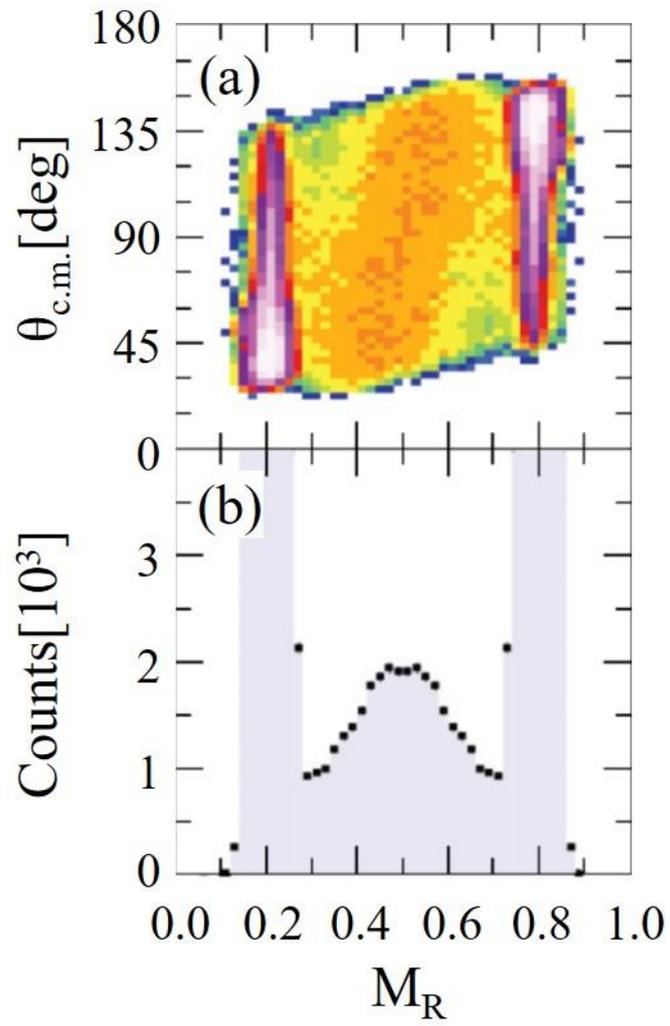

Fig. 3 for manuscript "FUSION PROCESS IN TRANSFER REACTION WITH..."

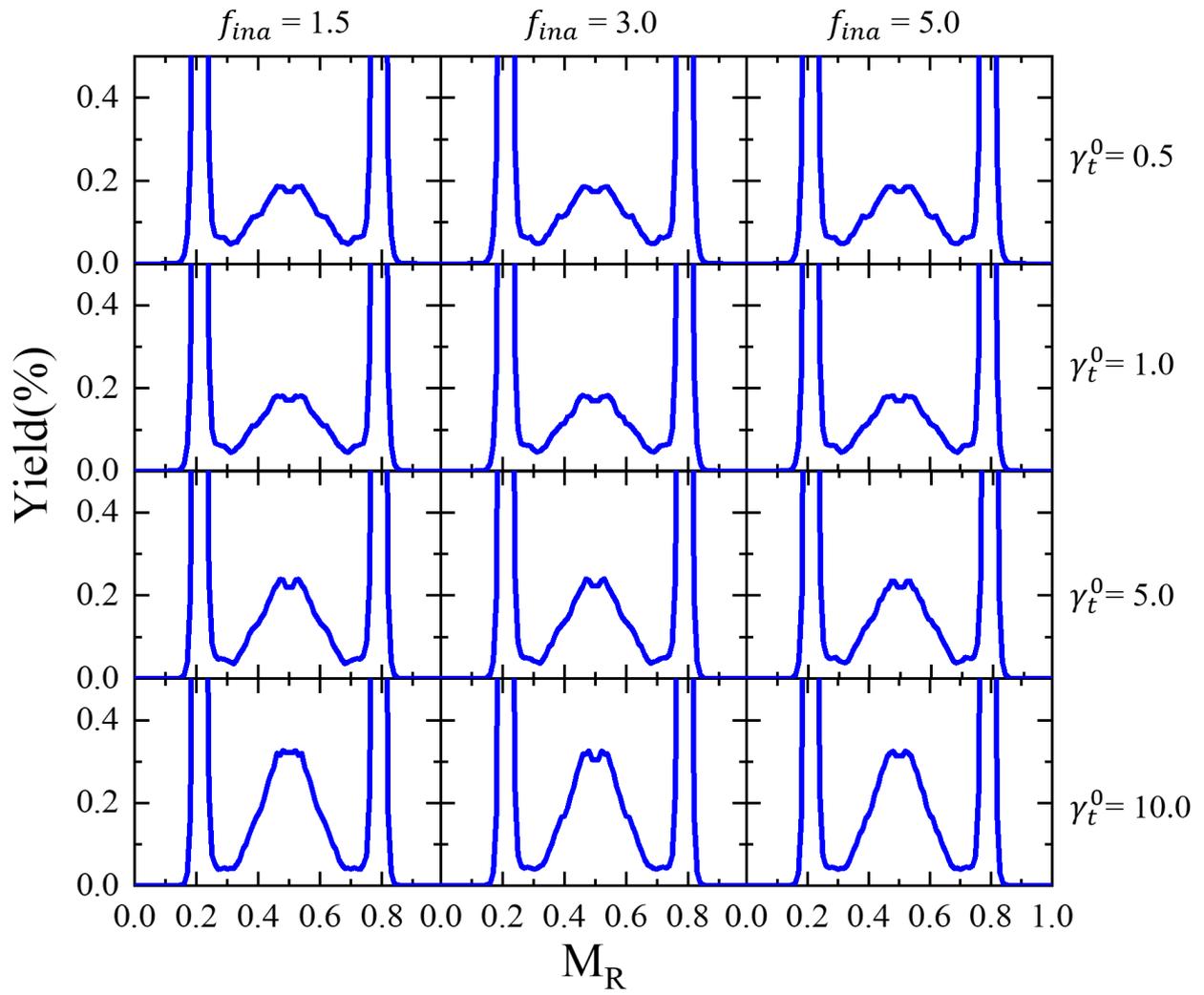